\documentclass[12pt]{article}

\usepackage{amsmath}
\usepackage{amsfonts}
\usepackage{amssymb}
\usepackage{graphics,graphicx,tikz}
\usepackage{subfig}
\usepackage{graphicx}
\usepackage{soul}
\usepackage{cite}
\usepackage{graphicx}
\usetikzlibrary{patterns}
\usepackage{enumitem}
\usepackage{array}
\usepackage{amsbsy}
\usepackage{bbold}\usepackage{appendix}
\usepackage{mathrsfs}
\usepackage{physics}
\usepackage{amsthm}
\usepackage{cancel}
\usepackage{yfonts}
\usepackage{inputenc}
\usepackage{chngcntr}

\numberwithin{equation}{section} %%
\usepackage[linktocpage]{hyperref}
\hypersetup{
	colorlinks=true,
	linkcolor=blue,
	filecolor=magenta,      
	urlcolor=blue,
	citecolor=blue
}
\def\beq{\begin{equation}}
\def\eeq{\end{equation}}
\newcommand{\commentOut}[1]{}
\def\eea{\end{align}}

\usepackage{color}

\begin{document}
\begin{titlepage}
%\hfill \hbox{CERN-PH-TH/2022-???}
%\vskip 0.1cm
%\hfill \hbox{NORDITA-2022-018}
\vskip 0.1cm
%\hfill \hbox{UPPSALA-19/22}
%\hfill \hbox{QMUL-PH-2?-??}
\vskip 1.5cm
\begin{flushright}
\end{flushright}
\vskip 1.0cm
\begin{center}
{\Large \bf Remembering Lars Brink and some of his work}

\vskip 1.0cm {\large  %Francesco Alessio$^{a,b}$, 
Paolo Di Vecchia
%$^{a, c}$
%, Carlo Heissenberg$^d$
} \\[0.7cm]

{\it \small  NORDITA, KTH Royal Institute of Technology and Stockholm University, \\
 Hannes Alfv{\'{e}}ns v{\"{a}}g 12, SE-11419 Stockholm, Sweden  }\\
 %{\it \small $^b$ Department of Physics and Astronomy, Uppsala University,\\ Box 516, SE-75120 Uppsala, Sweden}

{\it and }

{\it \small  The Niels Bohr Institute, Blegdamsvej 17, DK-2100 Copenhagen, Denmark}\\

%{\it \small $^d$ Queen Mary University of London, School of Mathematical Sciences, Mile End Road, E1 4NS, United Kingdom}\\

\end{center}
\begin{abstract}
In the first part of this paper I will  describe my work together with Lars and in the second part I will give a look at some of Lars's oldest papers. 
\end{abstract}
\vspace{1cm}
\begin{center}
To appear in the memorial volume for Lars Brink
\end{center}
\end{titlepage}

\tableofcontents

\section{Lars and me}
 \setcounter{equation}{0}

I first met Lars at Cern in 1972. We were part of a community of young people who were very excited to work on the Dual Resonance Model that later was shown  to be string theory.  
I remember having lunch almost every day with Lars, Eug{\`{e}}ne Cremmer and Joel Scherk. All of them are missed.
In 1973 Lars moved back to G{\"{o}}teborg  and in the fall of 1973 I visited him in G{\"{o}}teborg  and gave a seminar.

In May 1974 I moved to Nordita in Copenhagen on an assistant professor position.
Nordita is  a Nordic Institute, created  in 1957 by the five Nordic countries Denmark, Finland, Iceland, Norway and Sweden, when the Theory Division of Cern, that originally was in Copenhagen, moved to Geneva.  At Nordita, at that time, we were strongly encouraged  to  be in  contact with theoretical physicists of the various Nordic Institutions by travelling and giving seminars and lectures and by inviting Nordic people to visit Nordita and do the same. It was very pleasant to feel to be a part of a large Nordic community rather than just of a single country.  We used this encouragement  to meet regularly with Lars: he coming to Copenhagen and me going to G{\"{o}}teborg.

%From Cern I visited several times Torino and during these visits I got to know Alessandro and the rest of the group in Torino working on the Dual Resonance Model formed at that time by Alessandro D'Adda, Riccardo D'Auria, Nando Gliozzi, Ernesto Napolitano and Stefano Sciuto.

At that time I was working with a large group of italians from Florence, Naples and Torino  and we were open to collaborate with anybody who was interested.  
 The motto of our collaboration was: collaborate and help each other rather than compete. Lars joined  our group in the last year of  our collaboration publishing  3 papers in 1976: two with 12 authors~\cite{Ademollo:1975an,Ademollo:1976wv} and one with the same 12 authors plus John  Schwarz~\cite{Ademollo:1976pp}. 
 
 It was already known that the $N$-tachyon amplitude of the open Neveu-Schwarz (NS) model could be written in the following form~\cite{Fairlie:1973jw,Brink:1975qk}:
 \begin{align}
 &  A_N= \int \frac{\prod_{i=1}^N d Z_i }{dV_{abc}}\langle 0 \prod_{i=1}^N e^{ik_i X(z_i, \theta_i)}|0\rangle =  \int \frac{\prod_{i=1}^N d\theta_i dz_i}{dV_{abc}} \prod_{i<j} (Z_i-Z_j)^{2\alpha' k_i k_j}\, ,
 \label{A1}
 \end{align}
where  the super-coordinate $X(z) = x(z) + \theta \psi (z)$ has a bosonic and a fermionic component,
\begin{align}
& Z_i -Z_j = z_i -z_j - \theta_i \theta_j~~~;~~~ dV_{abc} = \frac{dZ_a dZ_b dZ_c}{[(Z_a-Z_b)(Z_a-Z_c)(Z_b-Z_c)]^{1/2}}\frac{1}{d \Theta}
\label{A2}
\end{align}
and $\Theta$ is the super-projective invariant variable: 
\begin{align}
& \Theta= \theta_c \Bigg[ \frac{(Z_a-Z_b)}{(Z_a-Z_c)(Z_b-Z_c)}\Bigg]^{1/2}- \theta_b  \Bigg[ \frac{(Z_a-Z_c)}{(Z_a-Z_b)(Z_b-Z_c)}\Bigg]^{1/2} \nonumber \\
& + \theta_a  \Bigg[ \frac{(Z_b-Z_c)}{(Z_a-Z_b)(Z_a-Z_c)}\Bigg]^{1/2} - \frac{\theta_a \theta_b \theta_c}{[ (Z_a-Z_b)(Z_a-Z_c)(Z_b-Z_c)]^{1/2}}\, .
\label{A3}
\end{align} 
If we compare it with the $N$ tachyon amplitude of the open bosonic string given by
\begin{equation}
A_N= \int \frac{\prod_{i=1}^N d z_i }{dV_{abc}}\langle 0 \prod_{i=1}^N e^{ik_i x(z_i )}|0\rangle =  \int \frac{\prod_{i=1}^N dz_i}{dV_{abc}} \prod_{i<j} (z_i-z_j)^{2\alpha' k_i k_j}
\label{A4}
\end{equation}
we see that, while in the open bosonic string each puncture is described by a single real variable $z$, in the open NS model is described by a bosonic variable $z$ and a fermionic variable $\theta$. This means that, while the bosonic string is invariant under the two-dimensional Virasoro  algebra including the $L_n$ generators, the NS string is invariant under the superconformal Virasoro algebra including the bosonic generators $L_n$ and the fermionic generators $G_{s}$. In \cite{Ademollo:1975an} we generalised this construction to the case of an $N$ extended world-sheet supersymmetry, corresponding of having $N$ fermionic variables $\theta_1, \theta_2 \dots \theta_N$ together with the bosonic variable $z$. 
In particular, for the case with $N=2$, we got a supersymmetric extension of the Virasoro algebra containing together with the Virasoro generators $L_n$ also two fermionic generators $G_s$ and ${\bar{G}}_s$ and an additional bosonic generators $T_n$, corresponding to a Kac-Moody $U(1)$ symmetry. 

In \cite{Ademollo:1976pp} we constructed the four-point amplitude for the lowest state that in this case is a massless scalar particle and we obtained the following amplitude:
\begin{equation}
A_4 \sim \frac{\Gamma (2-\alpha (s)) \Gamma (2-\alpha (t))}{\Gamma (2-\alpha (s) - \alpha (t))}~~;~~ \alpha (s) =1+  \alpha' s
\label{A5}
\end{equation}
that generalises the four-tachyon (with mass $\alpha'M^2=-1$)  amplitude of the bosonic string:
\begin{equation}
A_4 \sim \frac{\Gamma (-\alpha (s)) \Gamma (-\alpha (t))}{\Gamma (-\alpha (s) - \alpha (t))}~~;~~ \alpha (s) = 1+\alpha' s
\label{A6}
\end{equation}
and the four-tachyon (with mass $\alpha'M^2=-\frac{1}{2}$) amplitude   of the NS model
\begin{equation}
A_4 \sim \frac{\Gamma (1-\alpha (s)) \Gamma (1-\alpha (t))}{\Gamma (1-\alpha (s) - \alpha (t))}~~;~~ \alpha (s) =1+ \alpha' s\,.
\label{A7}
\end{equation}   
The hope was to find a consistent model with critical dimension $D=4$, but we found instead critical dimension $D=2$ . In \cite{Ademollo:1976wv} we constructed the string theory corresponding to $N=4$ getting no tachyon but negative critical dimension $D=-2$. Finally we constructed the four-point amplitude for the lowest state for the general $N$ case and we found the following amplitude~\cite{Ademollo:1976wv,Bruce:1976ng}:
 \begin{equation}
A_4 \sim \frac{\Gamma (N-\alpha (s)) \Gamma (N -\alpha (t))}{\Gamma (N-\alpha (s) - \alpha (t))}~~;~~ \alpha (s) = 1+\alpha' s\, .
\label{A8}
\end{equation}
This model has the external particles with mass $\alpha' M^2 = \frac{N}{2} -1$, no tachyon for $N \geq2$ but ghosts for $N>2$ for any $D>0$.

The amplitude in \eqref{A5} has in common with the four-point amplitude of the Lovelace-Shapiro model~\cite{Lovelace:1968kjy,Shapiro:1969km} that we write here
\begin{equation}
A_4 \sim \frac{\Gamma (1-\alpha (s)) \Gamma (1-\alpha (t))}{\Gamma (1-\alpha (s) - \alpha (t))}~~;~~ \alpha (s) =\frac{1}{2} +  \alpha' s
\label{A9}
\end{equation}
 the important property of having Adler zeroes, but, while the last amplitude has critical dimension $D=4$, the amplitude in \eqref{A5} has critical dimension $D=2$.  
 
 To be more specific,
 the amplitude in \eqref{A5} has a pole for $\alpha's =1$ corresponding to the exchange of a spin $2$ with positive norm and of a spin $0$, with residue  proportional to $ \frac{1}{D-1}-1$,
 that has zero norm for $D=2$ and is a ghost for $D>2$, while the amplitude in \eqref{A9} has a pole for $\alpha' s= \frac{3}{2}$ corresponding to the exchange of a spin $2$ and a spin $1$ with positive norm and of a spin $0$, with residue proportional to $ \frac{3}{D-1}-1$, that has zero norm for $D=4$ and is a ghost for $D>4$.    
  It turns out that   the amplitude in \eqref{A5} is the same four-point amplitude constructed in Ref.~\cite{Dong:2024klq} where the external pions  are obtained by a dimensional reduction of higher dimensional gluons~\footnote{I thank the authors of \cite{Dong:2024klq} for many very useful email exchanges.}.  In conclusion, the model with ${\cal{N}}=2$ world-sheet supersymmetry has only one particle that interacts according  to the two-dimensional non-linear $\sigma$-model, while the Lovelace-Shapiro model  provides a  consistent four-point amplitude~\cite{Veneziano:2017cks}, but its extension to $N>4$ external legs contains ghosts~\cite{Bianchi:2020cfc}. In other words, a consistent string extension of the non-linear $\sigma$-model in $D=4$ is not known and may be  impossible to construct with linearly rising Regge trajectory~\cite{Caron-Huot:2016icg}. 
 
 Using the fact that the $N=2$ string is described by two bosonic  and two fermionic coordinates, it can be seen as a four-dimensional field theory with two time and two space components~\cite{DAdda:1987ltc}. This model has been taken up again later in \cite{Ooguri:1991fp,Marcus:1992xt,Ooguri:1995cp} (see also the nice review by Marcus~\cite{Marcus:1992wi}). 
 
 In spring 1976 together with Lars and Paul Howe (at that time post-doc in Copenhagen)  we were trying to generalise the Nambu-Goto action to the superstring. We knew that this problem was not easy to solve because we had to construct the transformations and the Lagrangian  invariant under local $N=1$  supersymmetry and at that time the Lagrangian for $N=1$ supergravity was not yet known.  As an exercise we wanted to construct the non-linear Lagrangian for the point-like particle with spin $\frac{1}{2}$ (the $\alpha' \rightarrow 0$ limit of the Ramond model). But also this problem was not easy to solve until  we heard a seminar by Stanley Deser in Copenhagen on the newly found supergravity~\cite{Deser:1976eh}. After the seminar we sat down with Stanley and we wrote immediately  the local supersymmetry transformations and the non-linear action left invariant by them. Then Stanley went to Cern and also Bruno Zumino joined the collaboration.   This is the paper where the so called Polyakov action first appeared. The local supersymmetric Lagrangian describing the massless spin $\frac{1}{2}$ particle is given by:
\begin{equation}
L= \frac{1}{2} \bigg( \frac{{\dot{x}}^2}{e}- i \psi {\dot{\psi}}- \frac{i}{e}\chi \psi \cdot{\dot{x}}\Bigg)\, ,
\label{A10}
\end{equation}
where $x^\mu (\tau)$ describes the position of the particle and the fermionic coordinate $\psi^\mu$ describes the spin of the particle through the relation $S^{\mu \nu} \sim \psi^\mu \psi^\nu$. The action constructed from  $L$ is invariant under the following locally supersymmetric transformations:
\begin{align}
\delta x^\mu = i \alpha(\tau) \psi^\mu~~;~~\delta e=i \alpha (\tau) \chi~~;~~\delta \chi= 2 {\dot{\alpha}}(\tau)~~;~~\delta \psi^\mu = \frac{\alpha (\tau)}{e}\bigg({\dot{x}}^\mu - \frac{i}{2}\chi \psi^\mu \bigg)  \, .
\label{A11}
\end{align}
A mass can be given to the spin $\frac{1}{2}$ particle by adding an additional fermionic coordinate $\psi_5$ and  mass dependent terms  arriving at the following Lagrangian~\cite{Brink:1976uf}:
\begin{equation}
L= \frac{1}{2} \bigg( \frac{{\dot{x}}^2}{e} + em^2- i (\psi {\dot{\psi}}- \psi_5 {\dot{\psi}}_5)- i\chi ( \frac{\psi \cdot{\dot{x}}}{e} -m\psi_5)\Bigg)\, .
\label{A12}
\end{equation}
The correspond action is invariant under the locally supersymmetric transformations in \eqref{A11} together with
\begin{equation}
\delta \psi_5= m \alpha (\tau) + \frac{i}{me} \alpha (\tau) \psi_5({\dot{\psi}}_5- \frac{1}{2}m \chi) \, .
\label{A13}
\end{equation}
We can use the equations of motion to eliminate the auxiliary einbein $e$ and the gravitino $\chi$ to get~\cite{Brink:1976uf}:
\begin{equation}
L= m \sqrt{\bigg( {\dot{x}}- \frac{i}{m}{\dot{\psi}}_5\psi \bigg)^2} - \frac{i}{2}\psi {\dot{\psi}} -\frac{i}{2}  \psi_5 {\dot{\psi}}_5 \, .
\label{A14}
\end{equation}
This Lagrangian was independently obtained   in \cite{Barducci:1976qu}.

 Then we decided to meet with Lars in Copenhagen in August 1976 to work out the non-linear Lagrangian
for superstring. Lars came and together with Paul Howe  we worked for one week sitting down in the same  office on three different desks constructing by trials and errors the non-linear Lagrangian  for superstring and the local transformations that left the correspondent action   invariant by checking step be step that it transformed as a total derivative under the local supersymmetry transformations.  In the same week we wrote also the paper~\cite{Brink:1976sc}  that we sent to Physics Letters and appeared in the same volume where  the analogous paper written by Deser and Zumino~\cite{Deser:1976rb} was also published.  We got the following locally supersymmetric Lagrangian for the superstring:
\begin{align}
& L = e\Bigg[ -\frac{1}{2} \partial_\alpha x \partial_\beta x g^{\mu \nu}-\frac{i}{2} {\bar{\psi}}\gamma_\alpha \partial^\alpha \psi + \frac{i}{2}{\bar{\chi}}_\alpha \gamma^\beta \gamma^\alpha \psi \partial_\beta x +\frac{1}{8}({\bar{\chi}}_\alpha \gamma^\beta \gamma^\alpha \psi)({\bar{\chi}}_\beta \psi)\Bigg]\, .
\label{A15}
\end{align}
After we finished the paper I told Lars and Paul that  unfortunately I could not  continue to work in string theory because I wanted to get a permanent job and I would not get one if I continued to work on string theory that, at that time, was very unpopular.

This was the last paper that I wrote with Lars, but I kept strong contacts with him  during the nineties when  Lars became a member of the Nordita Board and soon after he was elected to be the chairman of the Board.
 He strongly contributed to make the Board more democratic and pushed to open Nordita to  new developments in theoretical physics (beyond traditional nuclear physics).

Last time I discussed physics with Lars was in occasion of the paper~\cite{Bianchi:2020cfc}  with  Bianchi and Consoli, on the extension of the Lovelace-Shapiro model to $N$ external legs for describing mesons. The reason was that  Lars was writing a paper~\cite{Brink:2019zdr}  that appeared in a memorial volume for Peter Freund where he was discussing the various attempts to find a consistent string theory describing the pseudo-scalar mesons. The modern way to formulate this problem is: can we construct a consistent string extension of the non-linear $\sigma$ model? There exist by now  several proposals~\cite{Carrasco:2016ldy,Mafra:2016mcc,Carrasco:2016ygv,Bianchi:2020cfc,Arkani-Hamed:2023swr,Dong:2024klq}, but all of them seem to have negative norm states (ghosts) because, unlike string theory, there are not  enough decoupling conditions  to eliminate them~\footnote{For a  discussion of the decoupling conditions see Ref.~\cite{DiVecchia:2007vd}.}.  In some case  ghosts appear already in the four-point amplitude, while, in other cases, only in higher point amplitudes.  It is probably not possible to write a consistent string extension of the non-linear $\sigma$ model with only linearly rising Regge trajectories~\cite{Caron-Huot:2016icg} and it would be interesting to use the bootstrap to clarify  this.

\section{A selection of early Lars papers}
 \setcounter{equation}{0}
 
 In this section I want to discuss some of the   papers that Lars wrote in the seventies and eighties. The first paper is the one where he and Holger compute the zero-point energy of the various string theories~\cite{Brink:1973oxp}. Since string theory contains an infinite number of harmonic oscillators it is expected that the zero-point energy  is divergent and, in order to extract a physical quantity from it, they introduce a cutoff  to regularise the expression. They then managed to extract  the zero-point energy by showing that it is independent on the explicit form of the cutoff. This quantity provides the mass of the lowest string state that, in the bosonic string,  turns out to be equal to
 \begin{equation}
m^2 = - \frac{D-2}{24 \alpha'}= - \frac{1}{\alpha'}~~~;~~ for~~~ D=26.
\label{SE1}
\end{equation}
The paper contains many physical considerations, but it is not easy to read. After the discovery of the $\zeta$-function regularisation, the zero-point energy for the bosonic string can be more easily computed in the following way~\cite{Gliozzi}:
\begin{equation}
\alpha' m^2 = (D-2) \sum_{n=1}^\infty n \Longrightarrow (D-2) \lim_{s\rightarrow -1} \zeta (s)  = -\frac{D-2}{24}\, ,
\label{SE2}
\end{equation}
where we have used the definition of the Riemann $\zeta$-function and its value for $s=-1$:
\begin{equation}
\zeta (s) = \sum_{n=1}^\infty n^{-s}~~;~~ \lim_{s\rightarrow -1} \zeta (s) = - \frac{1}{24}\, .
\label{SE3}
\end{equation}
 The next paper of Lars that I want to describe is the one~\cite{Brink:1973qm}  with David Olive where they construct the operator that projects into physical states. Lorentz covariance requires to write  the scattering amplitude  in terms of an infinite set of harmonic oscillators satisfying the algebra:
 \begin{equation}
[a_{n \mu}, a^\dagger_{m \nu} ]= \delta_{nm} \eta_{\mu \nu}
\label{SE4}
\end{equation}
that span a space that contains states  with an odd number of time components that have negative norm. Absence of ghosts requires, on the other hand,  that the physical space has positive definite norm. This was shown~\cite{Goddard:1972iy,Brower:1972wj} to happen for $D\leq 26$ where $D$ is the number of space-time dimensions. In particular, it was shown that the transverse states, called also DDF states~\cite{DelGiudice:1971yjh}, are complete for $D=26$. In \cite{Brink:1973qm} David and Lars construct an operator that projects into the space of physical states and they show  that, for $D=26$, the residue of each  pole of the amplitude is  factorised only in terms of the transverse states. 

The scattering amplitude involving  $M+R$  physical states of the bosonic string can  be written as follows:
\begin{equation}
{\cal{M}}_{M+R} = {}_M\langle p | D | p\rangle_{R}~~;~~ D = \int_0^1 dx x^{L_0-2}~~;~~ L_0 = \alpha' {\hat{p}}^2 + H~~;~~H=\sum_{n=1}^\infty n a_{n}^\dagger \cdot a_n\, ,
\label{SE5}
\end{equation}
 where $D$ is the string propagator, 
 \begin{align}
& {}_M\langle p| =  \langle 0, p_1| V(p_2, 1) D V(p_3,1) D \dots  V(p_M;1)  \nonumber \\
& |p\rangle_{R} = V (p_{M+1},1) D V(p_{M+2},1)D  \dots V (p_{M+R-1},1) |0, p_{M+R} \rangle  
\label{SE6}
\end{align}
 and $V(p_i,1)$ is the vertex operator of any physical state computed at the value of the Koba-Nielsen variable $z=1$.  As shown by Virasoro~\cite{Virasoro:1969zu}, the above string states satisfy the infinite set of conditions:
 \begin{equation}
(L_n - L_0 - (n-1))  |p\rangle_{N}=0~~;~~{}_M\langle p | (L_{-n} - L_0 -(n-1))=0~~;~~n=1,2 \dots \, .
\label{SE7}
\end{equation}
The residue at the pole with squared mass, given by $1-\alpha' q_N^2=N$
where $q_N =\sum_{i=1}^M p_i= -\sum_{i=M+1}^{M+R}p_i$, is given by:
\begin{equation}
{}_M\langle p |  \oint_0 \frac{dx}{2\pi i x}  x^{L_0-1} | p\rangle_{R}\, .
\label{SE8}
\end{equation}
On the other hand the projector into the space of transverse states is given by:
\begin{equation}
{\cal{J}}(k) = \oint_0 \frac{dx}{2\pi i} x^{{\cal{L}}_0 -H -1}~~;~~{\cal{L}}_0= \sum_{i=1}^{D-2} \sum_{n=1}^\infty A_{-n ;i} A_{n;i} \, ,
\label{SE9}
\end{equation}
where $A_{-n;i}, A_{n;i}$ are the creation and annihilation DDF harmonic oscillators~\cite{DelGiudice:1971yjh}.
If the transverse states are complete for $D=26$ we have to show the following identity:
\begin{equation}
{}_M\langle p | {\cal{M}}_N ( {\cal{J}}-1) | p\rangle_{R}=0~~~;~~~{\cal{M}}_N= \oint_0 \frac{dx}{2\pi i x} x^{L_0-1}\, .
\label{SE10}
\end{equation}
 In order to show this, they first show the validity of 
 \begin{equation}
{\cal{L}}_0-H\equiv E= (D_0-1)(L_0-1) + \sum_{n=1}^\infty (D_{-n}L_n + L_{-n} D_n)\, ,
\label{SE11}
\end{equation}
where
\begin{equation}
L_n = - \frac{1}{4\alpha'} \oint_0 \frac{dz}{2\pi i} z^{n+1} \left( \frac{\partial X}{\partial z}\right)^2~~;~~ D_n = \frac{1}{\sqrt{2\alpha'}} \oint_0 \frac{dz}{2\pi i} z^{n+1} \frac{1}{(k \frac{\partial X}{\partial z})} 
\label{SE12}
\end{equation}
and
\begin{equation}
X^\mu = {\hat{q}}^\mu +2\alpha' {\hat{p}}^\mu + i \sqrt{2\alpha'} \sum_{n=1}^\infty  \frac{1}{\sqrt{n}}( a_nz^{-n} - a_n^\dagger z^n) \, .
\label{SE13}
\end{equation}
They also show  that $L_n$ and $D_n$ satisfy the commutation relations
\begin{align}
& [L_n , L_m]= (n-m) L_{n+m} + \frac{D}{12} n(n^2-1) \delta_{n+m;0} \nonumber \\
& [L_n, D_m]= -(2n+m) \nonumber \\
& [D_n, D_m]=0 \, .
\label{SE14}
\end{align}
They imply
\begin{align}
& [L_n, {\cal{L}}_0 -H]= -n L_n \nonumber \\
& [L_n ,E]= -nL_n + \frac{D-26}{12} n(n^2-1) D_n
\label{SE15}
\end{align} 
 that are consistent with \eqref{SE11} for $D=26$. Three other relations, needed to show \eqref{SE10},  are
 \begin{equation}
X_n y^E = y^{E-n} X_n~~~;~~~X_n x^{L_0}= x^{L_0-n} X_n 
\label{SE16}
\end{equation}
for $X_n = L_n, D_n$, together with
\begin{equation}
(L_0+n-1)D_n =D_n(L_0-1) \, .
\label{SE17}
\end{equation}
Since $E\leq 0$ then 
\begin{equation}
X_n {\cal{J}}(k)=0 \, .
\label{SE18}
\end{equation}
We have now all that  we need to show \eqref{SE10} or equivalently  the following relation:
\begin{equation}
{}_M \langle p| \oint_0 \frac{dx}{2\pi i x}x^{L_0 - 1} (y^E-1) |p\rangle_R~~;~~y^E-1 = E\int_1^E dz z^{E-1}
\label{SE19}
\end{equation}
with $E$  given in \eqref{SE11} where we can neglect the first term with $L_0-1$ because ${\cal{M}}_N (L_0-1)=0$. Then we need to compute:
\begin{align}
&{ }_M \langle p| \oint_0 \frac{dx}{2\pi i x}x^{L_0 - 1} \sum_{n=1}^\infty \bigg( D_{-n} L_n+ L_{-n}D_n\bigg) \int_1^E  dz z^{E-1}  |p\rangle_R \nonumber \\
&= \sum_{n=1}^\infty { }_M \langle p| \Bigg[ (L_0+n-1) \oint_0 \frac{dx}{2\pi i x}x^{L_0 - 1-n}  D_n \int_1^E  dz\, z^{E-1}  \nonumber \\
&+ \oint_0 \frac{dx}{2\pi i x}x^{L_0 - 1} D_{-n} \int_1^\infty  dz z^{E-1} 
(L_0+n-1) \Bigg]|p\rangle_R  \, ,
\label{SE20}
\end{align} 
where we have used \eqref{SE16} and \eqref{SE7}.  Using Eqs. \eqref{SE16} and \eqref{SE17} we can write the previous expression as follows:
\begin{align}
& \sum_{n=1}^\infty { }_M \langle p| \Bigg[  D_n \oint_0 \frac{dx}{2\pi i x}x^{L_0 - 1} (L_0-1) \int_1^E  dz\, z^{E-1}  \nonumber \\
&+ \oint_0 \frac{dx}{2\pi i x}x^{L_0 - 1} (L_0-1) D_{-n} \int_1^\infty  dz z^{E-1} 
 \Bigg]|p\rangle_R =0\, ,
\label{SE21}
\end{align}
after using  the identities  $[L_0,E]=0$  and  ${\cal{M}}_N (L_0-1)=0$.

In conclusion, David and Lars showed  that, for $D=26$, the residue of each pole of the amplitude ${\cal{M}}_{M+R}$ in \eqref{SE5}  is saturated by the contribution of the transverse states. They are the only physical states and, since they have positive definite norm, the bosonic string is free of ghosts. 

In a subsequent paper~\cite{Brink:1973gi} David and Lars use the  previously constructed physical state projection operator to compute  the correct one-loop partition function subtracting the contribution of the unphysical states. They used the Feynman tree theorem. I remember that Lars was very happy  receiving a letter from  Feynman with congratulations about their paper.  They confirm the conjecture that the partition function, projected in the physical subspace, contains two less powers with respect to the unprojected one.

The final paper of Lars that I want to mention is the one in collaboration with Michael Green and John Schwarz~\cite{Green:1983hw} where they compute one-loop diagrams in the open and closed superstring and, by studying them  in the limit that the radii of the compactified dimensions and the Regge slope $\alpha'$ simultaneously approach zero, they obtain the one-loop scattering amplitudes,  with four massless particles, for ${\cal{N}}=4$ super Yang-Mills and for ${\cal{N}}=8$ supergravity.   The string amplitudes do not have  ultraviolet divergences. However, going to the case of four uncompactified dimensions ($D=4$), they find infrared divergences that are the same as those found by Weinberg~\cite{Weinberg:1965nx}  in 1965 in the limit where the external particles are massless. 

They start from the amplitude of four massless particles in type II superstring theory that is given by:
\begin{align}
& (Kinematical \,\,\, factor) \Bigg[ \frac{1}{stu} \frac{\Gamma (1-\frac{1}{2} \alpha' s ) \Gamma(1-\frac{1}{2} \alpha' t)\Gamma (1-\frac{1}{2} \alpha' u)}{\Gamma(1+\frac{1}{2} \alpha' s)\Gamma(1+\frac{1}{2} \alpha' t) \Gamma(1+\frac{1}{2} \alpha' u)} + c_1 g^{(1)}+\dots \Bigg]\, ,
\label{SE21a}
\end{align}
where the first term is the tree diagram and the second one is the one-loop contribution. $c_1$ is a numerical constant fixed by unitarity and 
\begin{equation}
 g^{(1)}=\frac{\kappa_{10}^2}{\alpha'} \int d^2 \tau  \tau _2^{-2} F(\tau) [F_2(a, \tau)]^{10-D}
\label{SE21b}
\end{equation}
with
\begin{align}
&F(\tau)=\tau_2^{-3}\int \prod_{i=1}^3 d^2 \nu_i\prod_{1\leq i <j \leq 4} \bigg( \exp[ -\frac{\pi (Im(\nu_{ji}))^2}{\tau_2}]|\Theta_1 (\nu_{ji};\tau)|\bigg)^{\alpha' p_ip_j}
\label{GSB2}
\end{align}
and
\begin{equation}
F_2 (a, \tau)= a \sqrt{\tau_2} \sum_{M,N} \exp\bigg(-2\pi i MN \tau_1-\pi \tau_2 (a^2M^2+\frac{N^2}{a^2}) \bigg)
\label{GSB22}
\end{equation}
is the toroidal compactification factor from ten to $D$ dimensions and $a= \frac{\sqrt{\alpha'}}{R}$. $\tau=\tau_1+i\tau_2$ is the parameter of the torus that is integrated in the fundamental region, while $\nu_i$ describe the punctures integrated in the following region:
\begin{align}
& 0 \leq Im \nu_i \leq \tau_2 ~~~;~~~~\frac{1}{2} \leq Re \nu_i \leq \frac{1}{2}
\label{GSB3}
\end{align}
with $\nu_{ji}=\nu_j-\nu_i$, $\nu_4=\tau$. The field theory limit is performed by sending simultaneously $\alpha'  $ and $R$ to zero and $\tau_2$ to infinity keeping the product $\alpha' \tau_2$ and $a$ fixed. In this limit the only relevant terms of the Green function in \eqref{GSB2} are the exponential factor and the $\sin$ factor contained in   $\Theta_1$.  After some calculation the field theory limit of $g^{(1)}$ is given by   
\begin{align}
& g_0^{(1)} \sim  \kappa_D^2 c(\gamma)\bigg( I_{\gamma} (s,t)+I_{\gamma} (t,s) + I_{\gamma} (s,u)+ I_{\gamma} (u,s)+ I_{\gamma} (t,u)+ I_{\gamma} (u,t)\bigg)\, ,
\label{SE22}
\end{align} 
where $\kappa_D^2$ is related to the $D$-dimensional Newton constant by  $\kappa_D^2= 8 \pi G_N$  and 
\begin{equation}
I_{\gamma} (s,t)= - \frac{t^{-\epsilon}}{tu}\int_0^1 dx \frac{(1-x)^{-1-\epsilon}}{x+ \frac{t}{u}}= - \frac{t^{-\epsilon}}{\epsilon t u}\Bigg[ -\frac{u}{ t}+ \int_0^1 dx \frac{(1-x)^{-\epsilon}}{(x+ \frac{t}{u})^2}
\Bigg] \,.
\label{SE23}
\end{equation}
The right-hand-side is obtained after a partial integration, 
 $\gamma = \frac{D}{2}-4= -\epsilon-2$ and $\epsilon= 2- \frac{D}{2}$  is the parameter that regularises the infrared divergences. 
 
 $\kappa^2_D$ is obtained from  $\kappa_{10}^2$  as follows.
   In the field theory limit $(F_2 (a, \tau_2))^{10-D} \sim a^{10-D} \tau_2^{5-\frac{D}{2}}$ as one can see from \eqref{GSB22}. This brings the total number of factors containing $\tau_2$ to be $d\tau_2 \tau_2^{-2} \tau_2^{5-\frac{D}{2}}$ (the factor $\tau_2^{-3}$ should not be counted because it disappears when we rescale  the variables $\nu_i$ to $\nu_i= \tau_2 \rho_i$). This means that, rescaling $\tau_2$ to get the field theory limit, we obtain an extra factor   $(\alpha')^{\frac{D}{2}-4}$ together with the factor $a^{10-D}$ from $F_2 (a, \tau)$ and they both  multiply  $\frac{\kappa_{10}^2}{\alpha'}$ obtaining
 \begin{equation}
\frac{\kappa_{10}^2}{\alpha'} a^{10-D}  (\alpha')^{\frac{D}{2}-4}= \kappa_{10}^2 R^{D-10}  =    \kappa_D^2  
\label{}
\end{equation}
 where, in the last step,  we have used the relation $\kappa_{10}^2= R^{10-D} \kappa_D^2$ between the $10$ and the $D$-dimensional Newton constant. This is the way that the factor $\kappa_D^2$ in \eqref{SE22} is obtained from the original factor $\frac{\kappa_{10}^2}{\alpha'}$ in \eqref{SE21b}.

 Going back to Eq.  \eqref{SE23} and using the following relation, valid up to linear order in $\epsilon$,
 \begin{align}
& \int_0^1 dx \frac{(1-x)^{-\epsilon}}{(x+ \frac{t}{u})^2}=-\frac{u^2}{s t}\bigg(1+ \epsilon \,\, {}_2F_1 (1,1,2;-\frac{u}{t})\bigg)= -\frac{u^2}{s t } -\epsilon  \frac{u}{s} \log (1 +\frac{u}{t})
\label{SE24}
\end{align}
we get
 \begin{align}
&I_\gamma (s,t)\sim -\frac{1}{\epsilon st}+\frac{1}{st} \log(-s)+ \dots
\label{SE25}
\end{align}
where $\dots$ indicate higher powers of $\epsilon$ that we do not need to consider. Inserting this expression in \eqref{SE22} we get:
\begin{align}
& g_0^{(1)} \sim \frac{\kappa_D^2}{\epsilon} \Bigg[ -\frac{2}{\epsilon}\bigg(\frac{1}{st}+ \frac{1}{su}+ \frac{1}{tu}\bigg)+ \frac{1}{st} \log(st)+  \frac{1}{su} \log(su)+ \frac{1}{tu} \log(tu) +\dots\Bigg]
\label{SE26}
\end{align} 
where the overall $\frac{1}{\epsilon}$ comes from $c(\gamma)$ given in Eq. (3.20) of their paper. The term proportional to $\frac{1}{\epsilon}$ inside the square bracket vanishes as a consequence of the equation $s+t+u=0$ valid for massless particles. The remaining quantity can be written as follows:
\begin{equation}
 g_0^{(1)} \sim \frac{\kappa_D^2}{\epsilon} \frac{1}{stu} \bigg(s\log s+t \log t+u\log u\bigg)= \frac{\kappa_D^2}{\epsilon stu} \bigg(\frac{s}{2} \log s^2+ \frac{t}{2} \log t^2 + \frac{u}{2}\log u^2-i\pi s\bigg)
\label{SE27}
\end{equation}
where we took into account that $s>0$ and $t,u<0$ in the physical region to get also the imaginary part.

In the case of four massless particles the infrared divergent factor, originally computed by Weinberg~\cite{Weinberg:1965nx}, can be found, for instance,  in Eq. (B.4) of Ref.~\cite{Alessio:2024wmz}
\begin{align}
& W_{phase} = \frac{G}{2\pi\epsilon}\Bigg( \sum_{j=1}^3 (-p_j k) \log \frac{4 (p_jk)^2}{m^2 \mu^2}  + 2i \pi    p_3k \Bigg)
\label{SE28}
\end{align}
where $\epsilon$ is related to the ratio of the infrared  and the ultraviolet cuts off, introduced by Weinberg, by $\frac{1}{2\epsilon}= \log \frac{\lambda}{\Lambda}$. 
Introducing the following Mandelstam invariants:
\begin{align}
& s= - 2p_1p_2=-2p_3k~~;~~t=-2p_1 k~~;~~u=-2p_2k
\label{SE29}
\end{align}
we can write $W_{phase}$ as follows:
\begin{align}
& W_{phase} = \frac{G}{2\pi \epsilon} \Bigg(  \frac{t}{2} \log \frac{t^2}{m^2 \mu^2}+  \frac{s}{2} \log \frac{s^2}{m^2 \mu^2}+  \frac{u}{2} \log \frac{u^2}{m^2 \mu^2} -i\pi s\Bigg)
\label{SE30}
\end{align}
We see that the amplitude, extracted from superstring in the field theory limit  and obtained in  \eqref{SE27}, is equal to the tree level amplitude in the field theory limit times the divergent Weinberg factor, apart from an overall numerical factor that we do not try to fix..

\subsection*{Acknowledgements}
I  thank Emil Bjerrum-Bohr,  Jin Dong,  Henriette Elvang, Henrik  Johansson,  Oliver Schlotterer,  Nabha Shah, and   Fei Teng  for many useful discussions and Nima Arkani-Hamed for email exchanges both on consistent string extensions of the non-linear $\sigma$-model.  I thank Raffaele Marotta for many  useful discussions on  infrared divergences in $D=4$  string theory and for a critical reading of the manuscript.     The research in this paper   is partially supported by the Knut and Alice Wallenberg Foundation under grant KAW 2018.0116.

\bibliographystyle{utphys}
\bibliography{hie4.bib}

\end{document}